# Divergence of effective mass in "Uncorrelated State Percolation" Model

Seyyed Mahdi Fazeli
*Department of Physics, Sharif University of Technology, Tehran, Iran*
Keivan Esfarjani
*Department of Physics, UC Santa Cruz, 95064 CA*

We want to answer the question of whether the divergence in the effective mass in metal-insulator transition (MIT) in 2DEG is in the same universality class as percolation. We use a model to make Percolated state in 2D and then calculate the effective mass in a supercell and the Bloch Theorem. It is seen that the effective mass, m*, scales as $m^* \propto (P - Pc)^{-\alpha}$ with $\alpha \approx 1$ and $P_c$ being the (classical) percolation threshold.

After the recent discovery of MIT in 2D electron (and hole) system by Kravchenko and his co-workers in 1994, (see references i &ii) people saw a divergence in spin susceptibility near the MIT (See references iii,iv,v,vi,vii&viii ) in the metallic phase. Spin susceptibility is proportional to effective mass and g-factor ( $\chi \propto g^* m^*$). Shashkin et al (in ref.ix) showed that effective mass, and not the g-factor, diverged near the MIT.

In 2002 Shi & Xie [x] proposed that Metal- Insulator Transition in 2DEG was appearing at or near the percolation transition. Later, Das Sarma & his co-workers[xi] and also others[xii] supported this view with measuring conductance near the MIT point. They showed that the conductance has a power-law dependence with the difference between the density (n) and a critical density ($n_c$),

$$\sigma(T) = A(T)(n - n_c(T))^{\delta(T)} \quad (1)$$

The critical density $n_c$, the exponent δ and the coefficient A all depend on the temperature of the sample.

Therefore they extrapolated $n_c(T)$ and $\delta(T)$ when temperature was decreased and found that: $\delta(T=0) \approx 1.4 \pm 0.1$. This formula is very similar to the conductance formula in 2D-percolation theory $\sigma \propto (P - P_c)^{\delta}$ with $\delta \cong 1.3$. (See references xi and xiii)

We use this view, and propose a non-interacting model for calculating the effective mass from the percolated ground state near the MIT point.

Our model is defined in a classical uncorrelated percolation network.

If we only use random hopping $H = \sum_{\langle i,j \rangle} t_{ij}\, C_i^+ C_j$ (2), according to the Anderson scaling theory of localization [xiv] it can be seen that all states are localized and they are not percolated. Hence effective mass is infinity, as is shown in figures (1).

To assess the localization of the eigenstates, we also calculate the inverse participation ratio (IPR) defined as:

$$IPR = \frac{\int |\psi|^4 dV}{\left(\int |\psi|^2 dV\right)^2} \quad (3)$$

For extended systems, it scales as the inverse of the volume of the system, and for localized states it is the inverse of the volume (area in 2D) spanned by the state:

$$IPR = \frac{1}{V_{effective}} \geq \frac{1}{L^2}$$

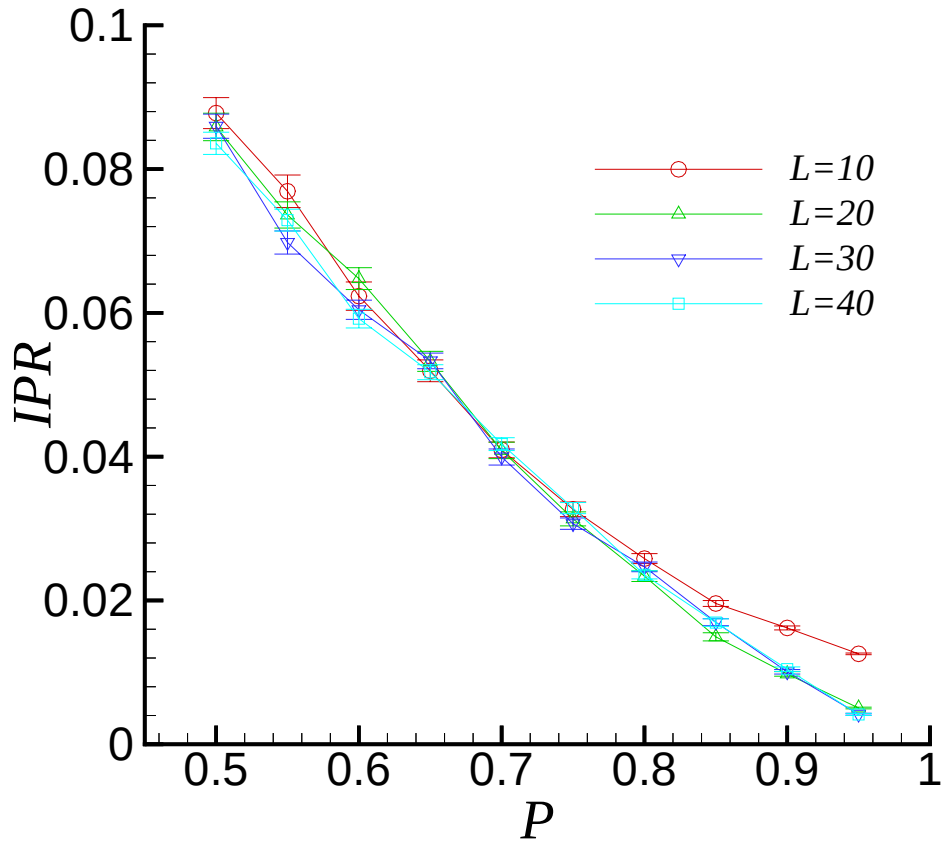

**Figure 1-ground state IPR versus probability of bond connection in percolation on different lengths using hopping Hamiltonian (see Eq.2).**

In Fig. 1 we show the IPR for the lowest eigenvector versus the probability of bond connection for the bond percolation network. Since the state is localized, the IPR value is the same regardless of the super-cell size. For P near 1, IPR curves split and tend to minimum conceivable IPR that is equal to $1/L^2$.

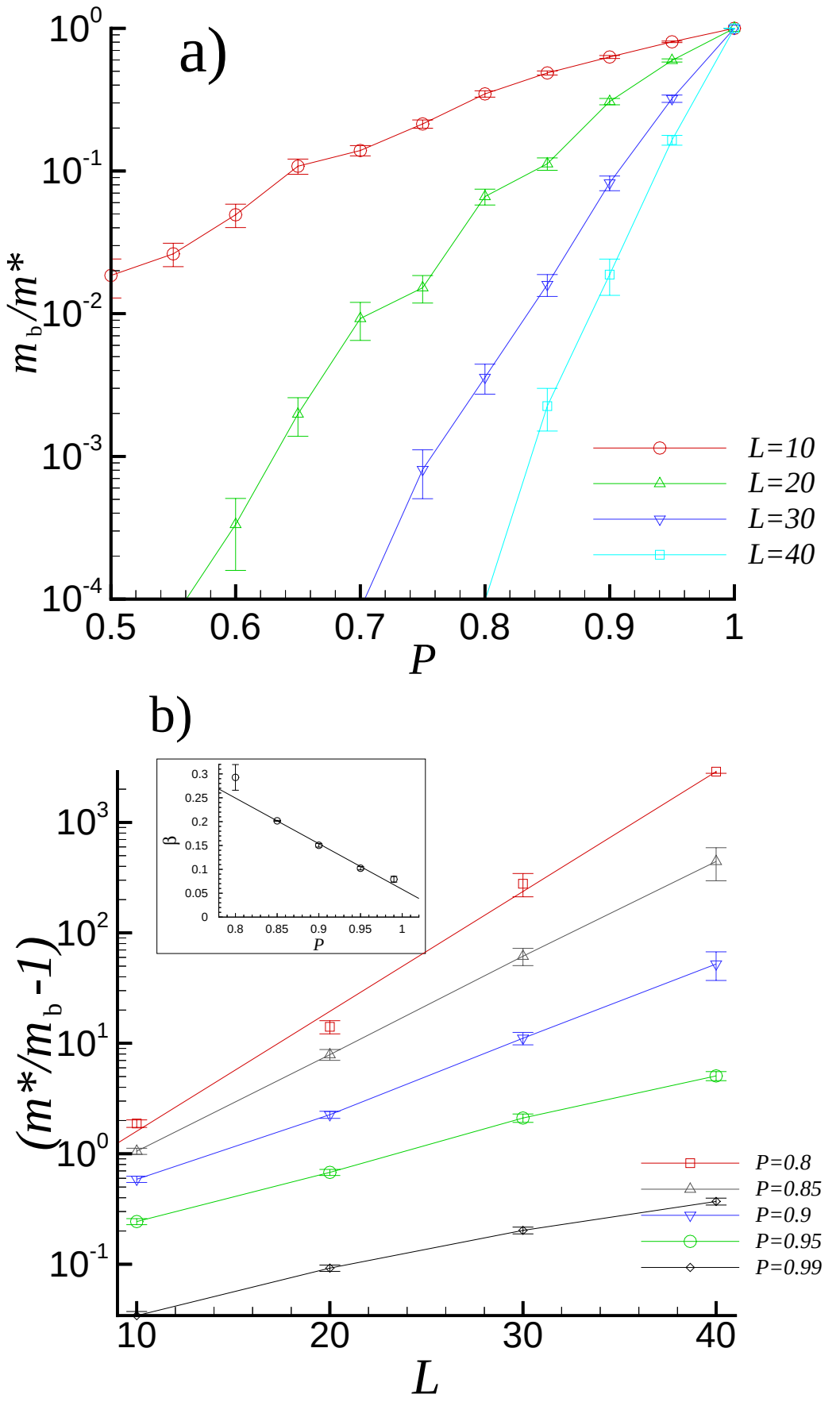

**Figure 2-a) $m_b/m^*$ versus probability of bond connection for different lengths. b) Plot ($m^*/m_b$-1) versus L, that behaves exponentially.** $(m^*/m_b - 1) \propto e^{\beta L}$, **Inset plots β vs P that shows** $\beta = 1.01 - 0.96P$ **.**

In figure 2 bulk mass ($m_b$ = m(P=1)) over the effective mass ($m_b/m^*$) versus P in the bond percolation is plotted. It can be seen that when the length of the percolation network (L) is increased, the effective mass increases rapidly (1/m* decreases) in any P<1. This means that at infinite L, the effective mass is infinite for any P<1.

IPR in figure 1 and $m_b/m^*$ in figure 2 are averaged over 100 different ensembles.

So we applied the onsite energy correlated with random hopping to make ground state extended and percolated. So our Hamiltonian is:

$$H = \sum \varepsilon_i \, C_i^+ C_i + \sum_{\langle i,j \rangle} t_{ij} \; C_i^+ C_j \quad \textbf{(4)}$$

Where $C_i^+$ is the creation operator on site issue i, $C_i$ is annihilation operator, $\mathbf{t_{ij}}$ is equal to $\mathbf{t_0}$ when site i is connected to site j with one bond, and equal to zero when they are not connected. Finally, $\varepsilon_i = -N \; t_0$ where N is the number of bonds connected to site i and $t_0$ is the hopping term that is negative. It is very simple to see that this Hamiltonian has an eigenvector of constant coefficients i.e. in coordinate space.

$\psi_0 = \sum_i C_i^+ \,|\, \Omega >$ & Ω is vacuum vector.)

In this model we use super cell with size L*L that is repeated in the square network. So we use the Bloch theorem to find eigenvalues of the Hamiltonian versus momentum (wave number) E(k) from which the effective mass can be deduced as:

$$\frac{1}{m_{effective}} = \nabla_k^2 E(k) \quad \textbf{(5)}$$

That is calculated in $\vec{k} = 0$ (The minimum energy of band structure). So we can approximate this formula to find effective mass with: $E(\vec{k}) \cong k^2 / (2 \, m_{effective})$ **(6)**

From Bloch theorem we have $\psi_{\vec{k}}(\vec{r} + \vec{R}) = \psi_{\vec{k}}(\vec{r}) e^{i \vec{k} \cdot \vec{R}}$ where L is a translation vector of the supercell. So we can find the effective Hamiltonian in the super cell whose form is similar to (Eq.4) the only difference being in the definition of hopping: $t_{lj}$ is zero when we have no bond between sites l & j, equal to $t_0$ when this bond is within one cell and equal to $t_0 \, e^{i \vec{k} \cdot \vec{R}_{lj}}$ when site issue l and site issue j are in two different cells with displacement vector $\vec{R}_{lj}$ between these two cells. ( $\vec{R} = \begin{Bmatrix} \pm L \, a \; \hat{x} \\ \pm L \, a \; \hat{y} \end{Bmatrix}$ and "a" is the length of one bond.)

When all bonds are connected it's easy to see that

$$E(\vec{k}) = 2 \, t_0 \left( Cos(k_x \, a) + Cos(k_y \, a) - 2 \right) \approx -t_0 \, a^2 \, \vec{k}^2$$

when k is very small. So $m_b$ (bulk) is equal to $-\dfrac{1}{2 \, t_0 \, a^2} (\hbar = 1)$.

In practice, $E(\vec{k})$ is calculated for 5 different small k-points near the zone center, and then a line if fitted to $E(\vec{k})$ versus $k^2$; the slope of this line is equal to $\dfrac{1}{2 \, m_{effective}}$. These values are averaged over 100 different network configurations. Finally, the effective mass is calculated for several lattice lengths L, within both bond and site percolation models.

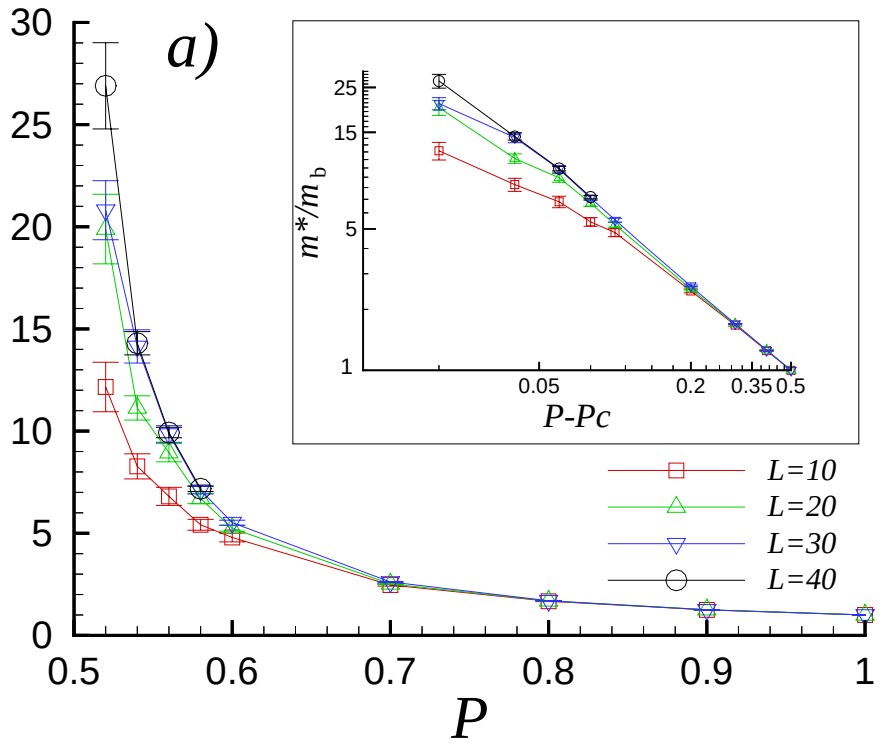

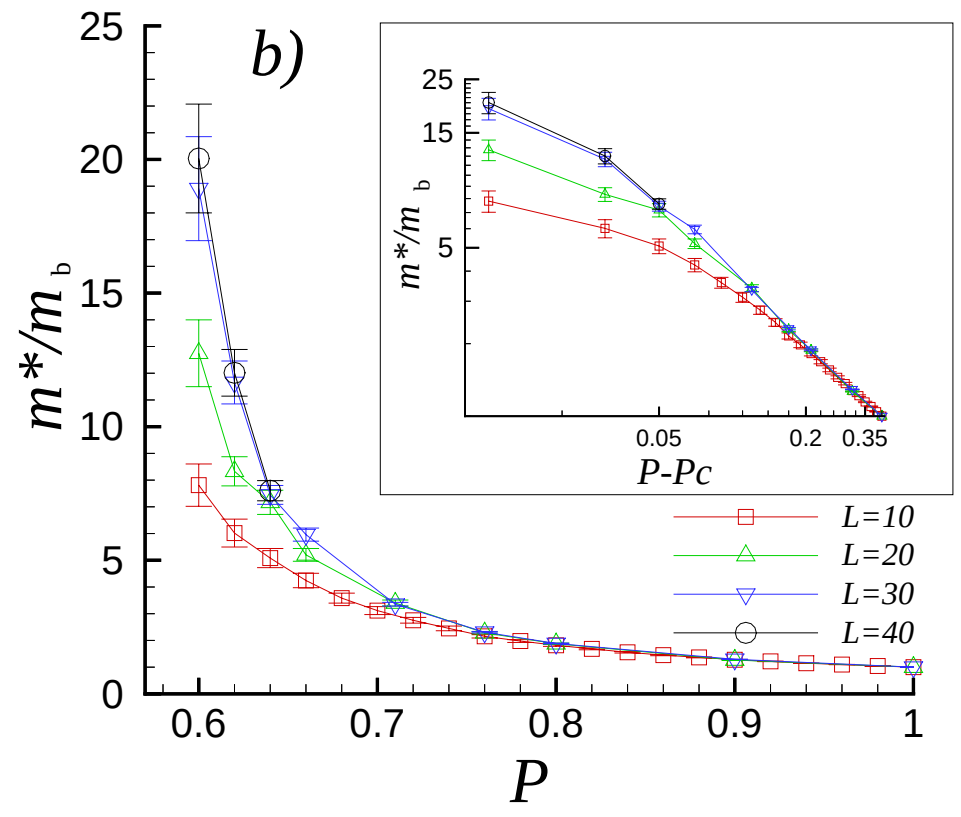

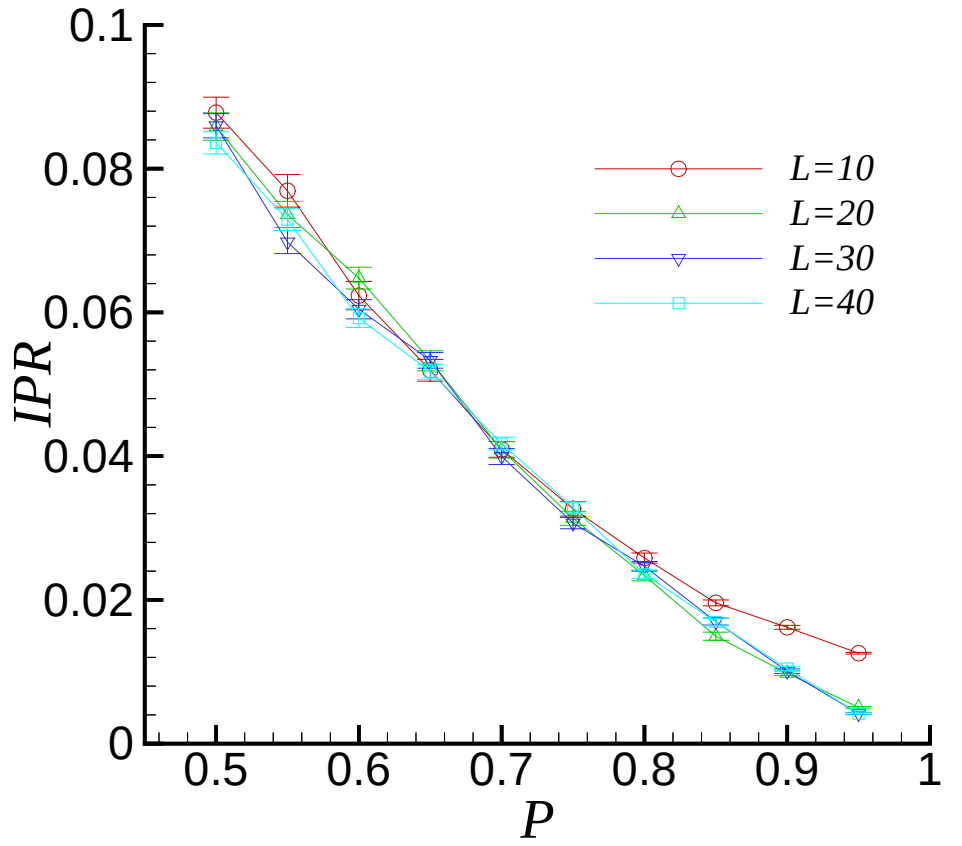

**Figure 3-Effective mass versus (P) (probability of bond connections in bond percolation (Fig. a.) and probability of site existence in site percolation (Fig. b.)). Inset: Log-Log Plot of effective mass versus (P-Pc) (Pc = 0.5 in bond percolation and Pc = 0.59 in site percolation (see ref. xiii)).**

In these two cases we can find that: $m^* \propto (P - Pc)^{-\alpha}$ .

α is equal to $1.00 \pm 0.05$ in bond percolation & equal to $0.9 \pm 0.1$ in site percolation. If we change site percolation threshold probability ($P_C$) from 0.59 to 0.58 we have a better fit and α becomes $1.00 \pm 0.05$ in site percolation.

For smaller length super cells we see finite size effects that underestimate the effective mass.

In all kinds of percolation models (correlated or uncorrelated in any dimension), we can see the divergence of effective mass near the percolation threshold, but the exponent (δ) and percolation threshold ($P_C$) will be different.

In experiments, it is seen that $m_{effective} \propto \chi_{effective} \propto (n - n_c)^{-\alpha}$ (g-Factor has limited value and more than zero at the MIT point.(see ref. ix)). Reported values of α by different groups are different:

$\alpha = 0.60 \pm 0.12$, α =0.24 & α = 0.27 (In ref v & vii), and α = 1 (In ref. iv & viii)

Das Sarma & his co workers (in ref. xi) **assume** that: $(n - n_c) \propto (P - P_c)$ and also **assume** MIT in 2DEG is **uncorrelated** percolation transition. If we accept these assumptions, our work is in good agreement with results of Shashkin, Kravchenko et al. (see references iv and viii).

In conclusion, we used a modified tight-binding Hamiltonian for making percolated states, and by using super cells and Bloch theorem, calculated the effective mass. We saw that it diverged near the percolation transition as $m^* \propto (P - Pc)^{-\alpha}$ .

We thank M. Vahedi, H. Cheraghchi & A. Saberi for their helpful discussions.

---